# On the Optimization of non-Dense Metabolic Networks in non-Equilibrium State Utilizing 2D-Lattice Simulation

## Abstract

Modeling and optimization of metabolic networks has been one of the hottest topics in computational systems biology within recent years. However, the complexity and uncertainty of these networks in addition to the lack of necessary data has resulted in more efforts to design and usage of more capable models which fit to realistic conditions. In this paper, instead of optimizing networks in equilibrium condition, the optimization of dynamic networks in non-equilibrium states including low number of molecules has been studied using a 2-D lattice simulation. A prototyped network has been simulated with such approach, and has been optimized using Swarm Particle Algorithm the results of which are presented in addition to the relevant plots.

**Introduction**

The objective of Systems Biology is to study the complex biological processes as integrated systems of many interacting components [14]. Current metabolic engineering processes allow manipulating metabolic networks to improve the desired characteristics of biochemical systems [1]. While researches and investigations are still ongoing on the subject of metabolic networks and genome scale modeling, two crucial approaches are frequently utilized. The first approach is to consider all the reactions in our favorite pathway to be occurred with fixed rate which usually happens when the whole system is in steady state. Therefore, optimizing the yield of desired product is strait forward as the problem is a linear programming with linear cost function and constraints. The second approach is dynamical modeling which allows reactions to occur with any non-linear rate which is indicative of a system in non-equilibrium system. Kinetic models describe the complete dynamics of the network, and have proven useful to implement optimization and control over the network, such as in [2]. The creation of reliable kinetic models involves the estimation of parameters, the complexity of this task increasing with the size of the network considered [1].

Optimizing a metabolic network is directly dependent on the network's model. Several strategies has been introduced in order to optimize the kinetics of networks some of which are [3] [4-12]. The majority of these approaches are based upon mixed-integer linear programming (MILP) that employ integer variables to capture the discrete nature of decision making required to analyze, curate and redesign metabolic networks. [13].

When optimizing a metabolic network for a given objective two different questions should be answered. The first is to find which branch or branches must be manipulated. The second is to determine what type of alterations must be done. Strategies such as OptKnock [8] and the work in[9] address the first problem. In this work a strategy for the second problem is described.

In this paper, we introduced 2-D lattice simulation for modeling dynamic and sparse metabolic networks and optimized this model utilizing Particle Swarm Optimization. In the following sections, firstly, the preliminary assumptions of 2-D lattice simulation are presented. In section 2, a prototyped network is introduced which has been used in many scientific works on the same subject. In section 3, we present the result of the optimization, and discuss the results.

**2-D Lattice Simulation**

Consider a 2-D lattice where each cell can be referred to a molecule or a blank space. Should a molecule is larger than the rest particles in a pathway, allocating more than one cell to that molecule is achievable with ease. Moreover, consider a neighborhood area in which a molecule can react with the other particles. Regularly, a neighborhood of one cell from each side of a cell can be considered as the sensitive area. However, this point is also adaptable to the type of the molecule in hand similar to its size. If two types of molecules which can react with each other are placed in their sensitive area in a 2-D lattice, then the reaction will take place in the way that those cells will convert to the products of the reaction they are involved. This process includes any kind of reactions such as those with enzymes or the reactions which have many substrates but much less products. Assuming that in each period of time just one reaction can take place, the algorithm of this simulation can be presented as follows:

1- Generate a 2-D lattice with the expected size.
2- Allocate each cell to a molecule or blank space according to the initial volume of each material.
3- Determine the total number of steps.
4- For each step pick a random cell from the lattice.
5- If a relevant molecule exists in the pre determined sensitive area of the chosen cell, substitute the cells with products with the probability of *P*.
6- Repeat steps 2-5.

Figure 1: samples of variant molecules in red and their sensitive area in blue in a 5*5 lattice.

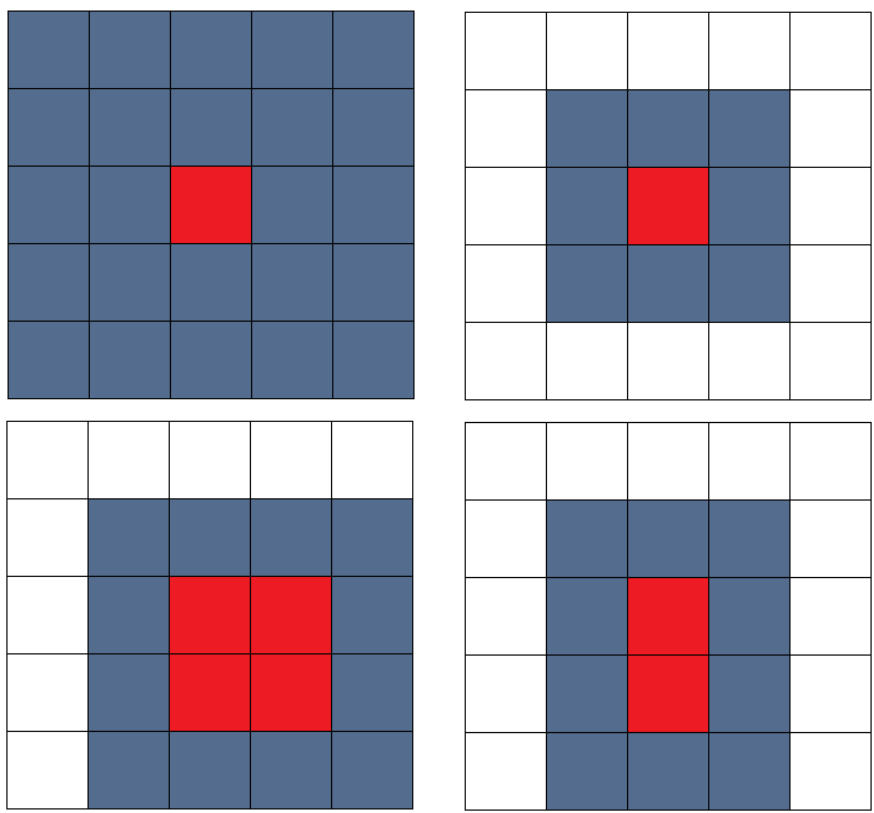

**Figure 2: Result of 2-D lattice simulation of equation (1): Change in concentrations over time for enzyme E, substrate S, complex ES and product P in a (100*100) lattice with 0% neutralized molecules.**

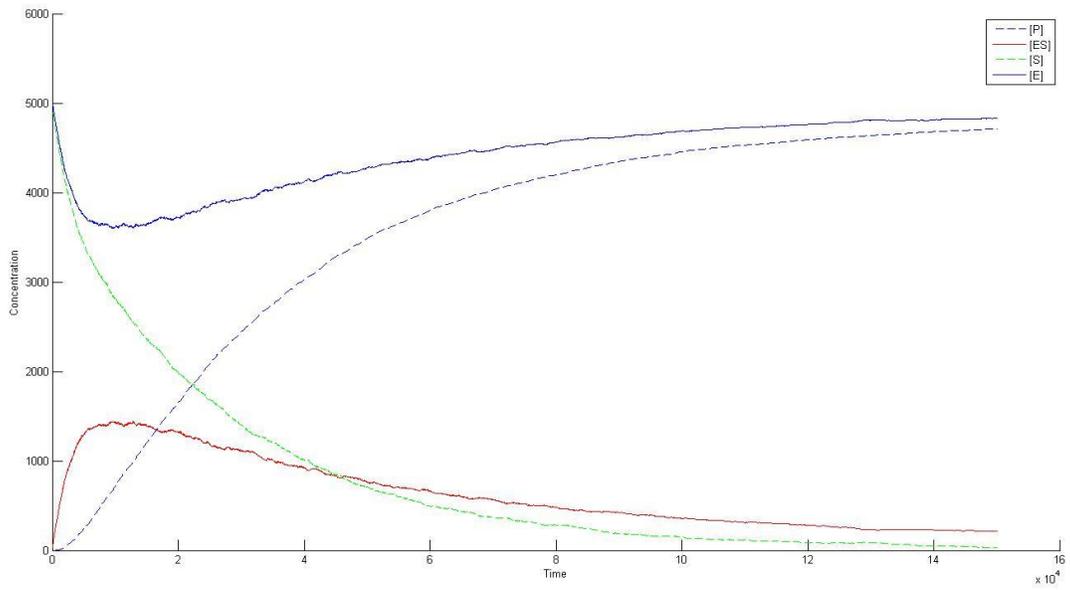

**Figure 3: Result of 2-D lattice simulation of equation (1): Change in concentrations over time for enzyme E, substrate S, complex ES and product P in a (100*100) lattice with 50% neutralized molecules.**

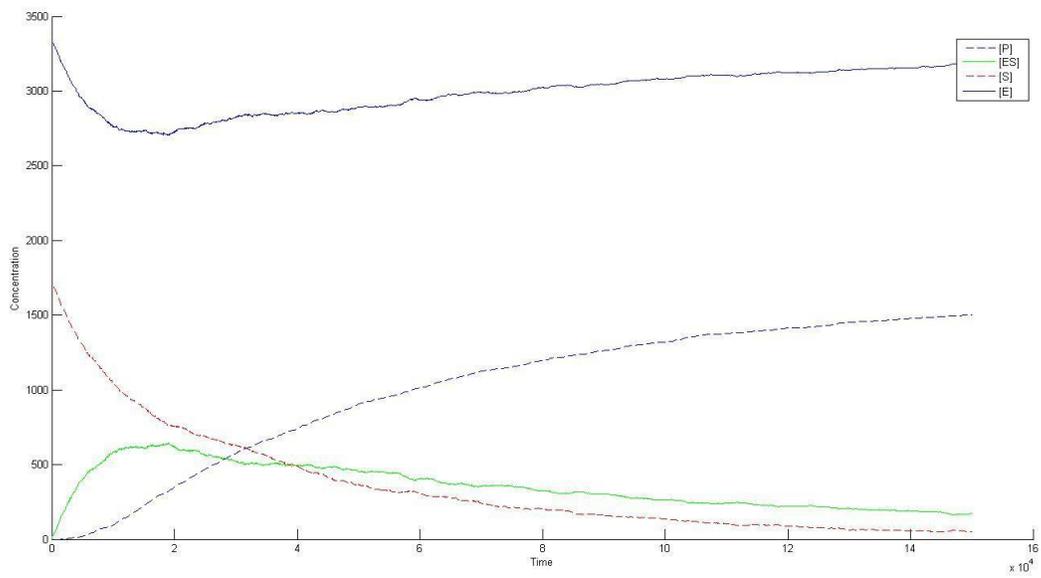

**Figure 4: Result of 2-D lattice simulation of equation (1): Change in concentrations over time for enzyme E, substrate S, complex ES and product P in a (100*100) lattice with 75% neutralized molecules.**

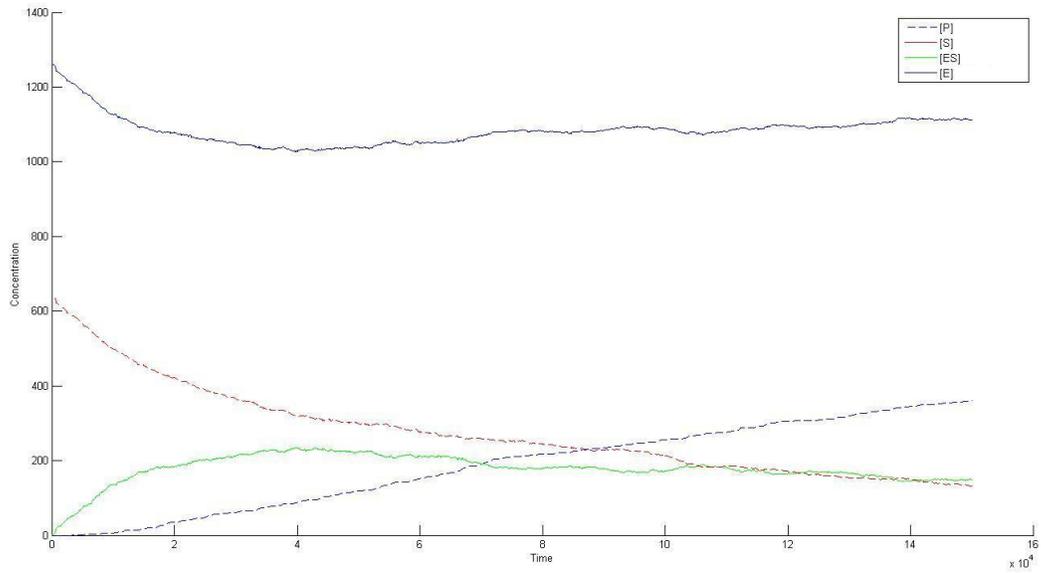

**Figure 5: Result of 2-D lattice simulation of equation (1): Change in concentrations over time for enzyme E, substrate S, complex ES and product P in a (100*100) lattice with 81% neutralized molecules.**

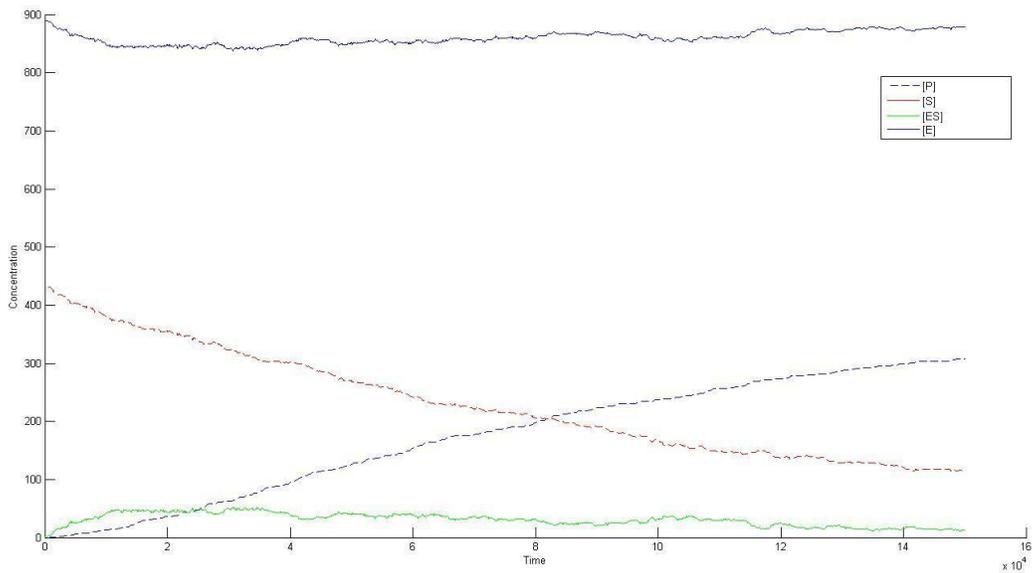

Before applying this method into a real life problem, we have examined with some simple, theoretical problems such as Michaelis-Menten equation. The results pretty match the solution of equation in the case of large number of molecules and in large scale 2-D lattice.

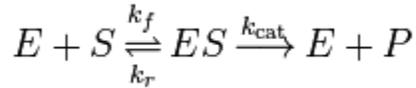

$$E + S \underset{k_r}{\overset{k_f}{\rightleftharpoons}} ES \overset{k_{cat}}{\longrightarrow} E + P$$

**Time Scaling**

As one time step in simulation is a short period, and its length is not clear, one should illustrate the length by realizing the duration of time that a pathway terminates in reality. For instance, should a pathway terminates after 30 seconds while its simulation after 12000 steps, it indicates that each time step has the length of 30/12000=0.0025ˢ.

**Prototype Network**

The prototyped network is the one tested in [BMC] which is itself a modified version of [16]. There are however, some changes in the figure due to different considerations in this paper. A graphical representation of the network is shown in Figure 2:

Figure 6: Prototype network. The circles correspond to metabolites and the arrows to fluxes with the reaction rates indicated. The figure is a regulated version of the figure presented in [1].

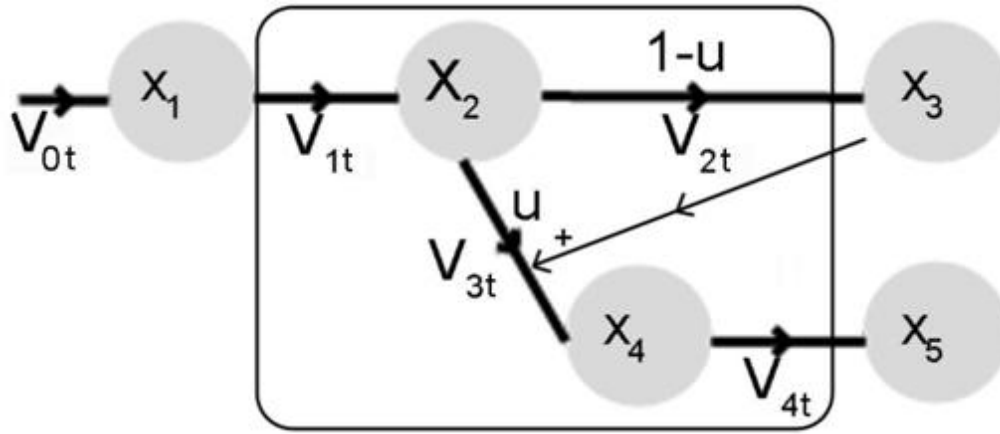

Here the states $x_i$, $i = 1,..., 5$ are metabolite concentrations at the network nodes, $v_{it}$, $i = 1,..., 4$ are fluxes associated to the metabolic network branches in time $t$. In the figure, $u$ represents a control function that allows redirecting the flux between the branches $x_2 \rightarrow x_3$ and $x_2 \rightarrow x_4$. Assuming that $x_3$ represents a precursor of the cellular objective (such as growth) and $x_5$ the desired product, if $u(t)$ is biased towards the branch of $v_2$ this yields the formation of $x_3$ but little or no production of $x_5$. If $u(t)$ is biased towards the branch of $v_3$ the production of $x_5$ will be affected by the low concentration of $x_3$ (since there is a forward feedback). Thus, there is an optimal profile for $u(t)$ to maximize the concentration of $x_5$ at the final time $t_{final}$.

**Optimization Problem**

The optimization problem associated with prototyped network is finding the time in which the pathway should be switched from reaction (2) to (3) so that the yield of product (5) will be at its maximum. While in the literature time has been always considered as an integer value, here we assumed it as a contentious value. Moreover, there are other assumptions which were considered in this paper:

1- $V_{0t}=0$ for $t \in [0 \ t_{final}]$,
2- The probability for a molecule of each substrate to be placed in a cell of the lattice after each time step is similar for all substrates.
3- Initial concentration of $X_0$ and the concentrations of all the enzymes related to each reaction are the same.
4- Each reaction needs just one enzyme.

Two different simulations were designed associated with the number of blank cells or neutralized molecules in the lattice, and each of them were run for 5 times. The lattice size was 100*100 for both simulations. The number of steps was supposed 100,000; however, this time should be regulated according to the problem in hand. The optimization algorithm utilized in this paper was Particle Swarm Algorithm with 10 number of swarms and 10 number of generations. The fitness of the SPO algorithm was considered $(1/[X_5])$. Since the nature of the simulation is probabilistic, each simulation had to be run for five times.

The results are in agreement with the fact that the denser the biological systems, the less time we demand to achieve the maximum yield of desired product. The Switching times for the lattice where 50% and 20% of the molecules are neutralized are 3.1628e+004 and 2.1088e+004 respectively.

Figure 7: The result of optimization process on the 100*100 lattice model with 0% of neutralized molecule after 10 generations.

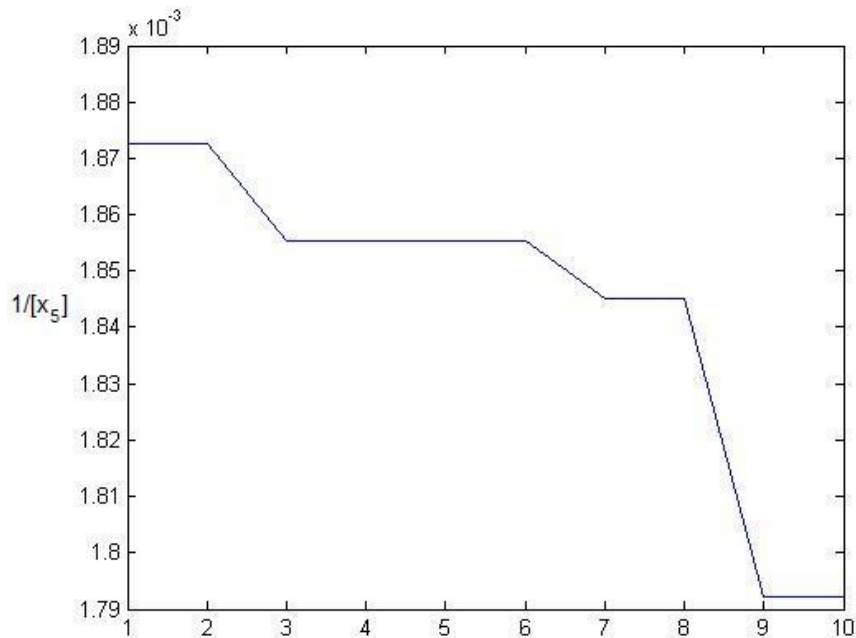

Figure 8: The result of optimization process on the 100*100 lattice model with 50% of neutralized molecules after 10 generation.

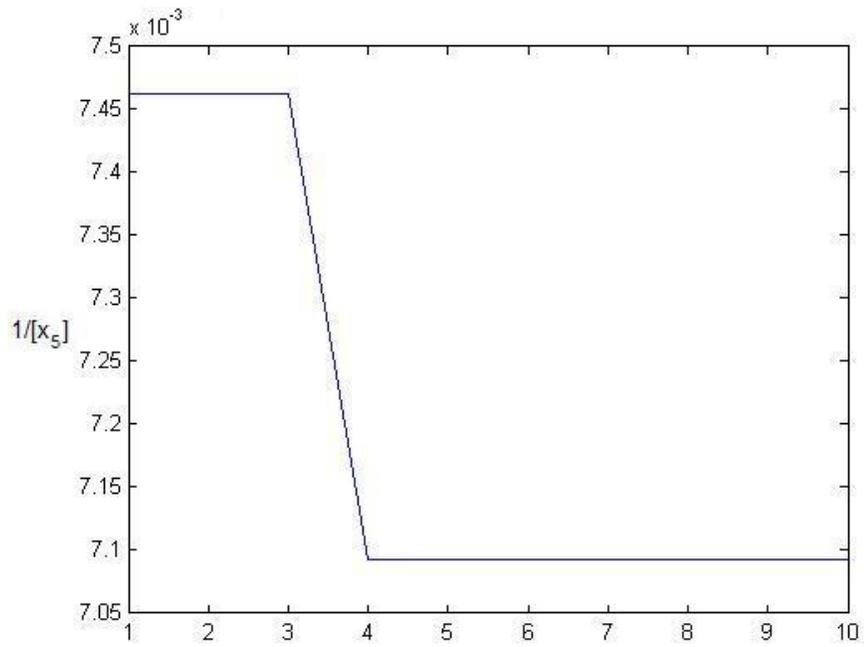

Figure 9: The Network dynamics where regulation time is equal to 2.1088e+004.

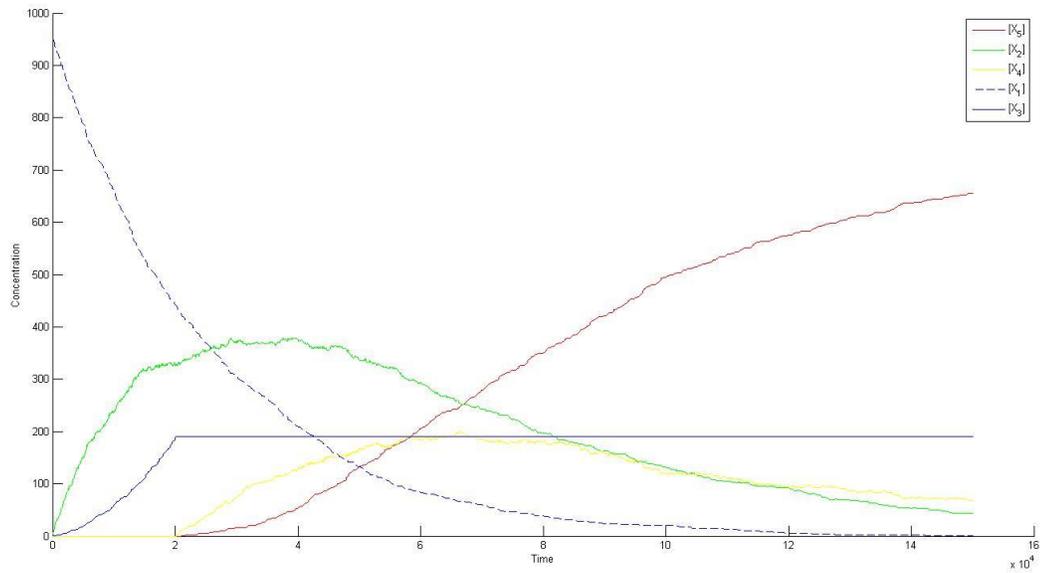

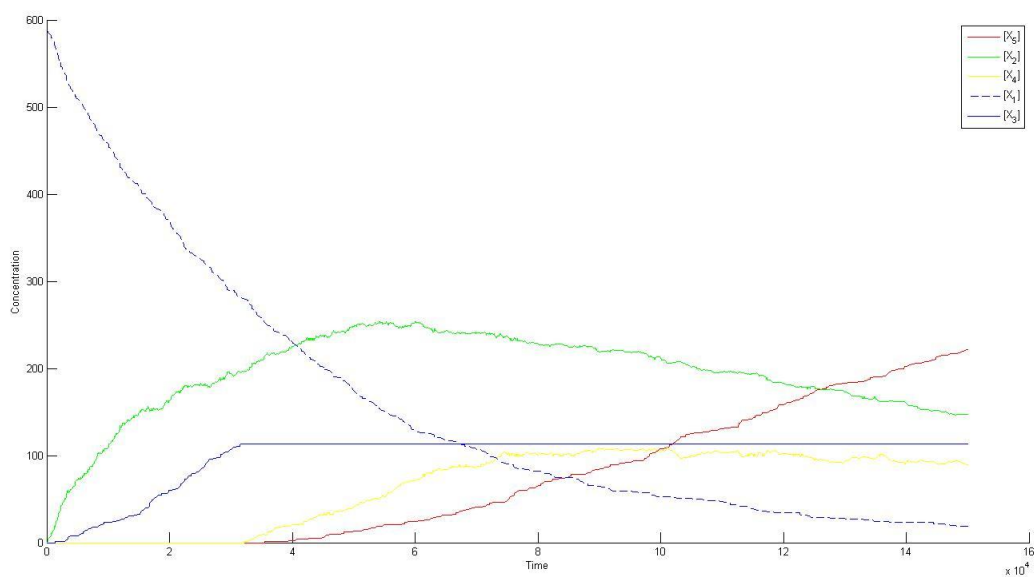

Figure 10: The Network dynamics where regulation time is equal to 3.1628e+004.

## Conclusion

2-D lattice simulation has been introduced as a tool to model the kinetics of metabolic networks in an efficient way. It has been shown that networks with low number of molecules can modeled, and optimized with heuristic algorithm (SPO).

It should be emphasized that the presented results are achieved based on the assumptions about the movement of molecules in each time step, the lattice size, and also the size of the molecules. In order to have more accurate achievements and better performance in special cases, those factors should be changed accordingly. These changes do not necessitate challenging efforts, and can be applied with ease.

The 2-D lattice simulation has also the capability of being modified according to more complex networks impacted by perturbations. Also, the equilibrium condition can also be simulated using this strategy all of which are to be presented in future works.